\documentclass[twocolumn]{article}
\usepackage{epsfig,cite,amsmath}
\begin{document}
\title{
Second order magnetic contributions to the\\
hyperfine splitting of the
{\it 5snd $^1\!$D$_2$} states in $^{87}$Sr.}
\author{J. G\"udde, A. Klinkm\"uller \footnote{G{\"o}teborgs universitet
  and Chalmers tekniska h{\"o}gskola AB, Avd.~f{\"o}r atomfysik,
  Fysikgr{\"a}nd 3, SE-412\,96 G{\"o}teborg, Sweden}, P. J. West, and E.
  Matthias\\ Institut f\"ur Experimentalphysik, FU Berlin, Arnimallee
  14, 1000 Berlin 33, Germany}
\date{20 January 1993}
\maketitle
\begin{abstract}
The hyperfine structure of $5snd\;^1\!D_2$ states in $^{87}$Sr was 
studied for principal quantum numbers $11\le n \le 25$ by Doppler-free 
two-photon excitation combined with thermionic detection. 
Based on the more accurate data, compared to previous results, the 
hyperfine structure was analysed by second order perturbation theory
using MQDT wavefunctions. As a result we find that the surprisingly
large $B$-factors obtained by fitting the data with the Casimir
formula do not reflect a quadrupolar interaction but rather consist 
predominantly of second order magnetic dipole contributions.
Further, inconsistencies in the King plot of odd-even isotope shifts 
were removed by this type of analysis. The result demonstrates the 
importance of higher order effects in hyperfine splittings of excited 
states in atoms with two valence electrons, even when the fine 
structure splitting is large compared to the hyperfine splitting.
\end{abstract}
\begin{flushleft}
{PACS numbers: 35.10.Fk, 32.80.Rm, 31.30.Gs}
\end{flushleft}
\section{Introduction}
Many problems in atomic physics can be treated using perturbation 
theory. Normally for low lying states with large energy distances to 
neighbouring levels the application of perturbation theory in first 
order gives satisfying results. 
However, the level spacing of high lying states (Rydberg states)
is considerably smaller and higher order perturbation theory 
is required for correct treatment.

The electron configuration of excited states of the alkaline earth 
elements is determined by the two electrons outside a core of closed 
electron shells.
The existence of low lying bound doubly excited states 
causes configuration mixing and influences the coupling of the two 
valence electrons (singlet-triplet mixing).
The multichannel quantum defect theory (MQDT) has proved to be 
an efficient tool \cite{Aym_84} for the description of such 
perturbed Rydberg states with only a small set of parameters.

Particularly the atomic hyperfine structure (hfs) is sensitive to 
admixtures in the wave function of Rydberg states \cite{MRB_83} and 
is therefore a most sensitive quantity for investigating such 
phenomena.

For the evaluation of the hfs it is in most cases sufficient to 
consider only the magnetic dipole and the electric quadrupole 
interaction. Then it is possible to parameterize the hfs -- splitting 
in first order perturbation theory by the hyperfine constants $A$ and
$B$, using the Casimir formula. The hyperfine constants $A$ and $B$  
measure the magnitude of the magnetic dipole and the electric 
quadrupole interaction, respectively.
This kind of evaluation and separation of the interactions assumes 
that contributions of higher order perturbation theory are negligible.
If that is not the case the hyperfine constants lose their physical 
meaning.

Previously, the effects of higher order contributions to the hfs in 
Rydberg states due to neighbouring fine structure levels have been 
observed in 
Hg\cite{ScJ_32,Cas_32,GoB_33},
$^3$He\cite{LFP_80},
Ca\cite{BeT_82_1},
Sr\cite{MRB_83,BeT_82_1,BMT_81_1,BeT_82_2,BST_82},
Ba\cite{MRB_83,BeT_82_1,GZJ_82,NeR_82_1,NeR_82_2,ElH_83,RNM_82},
Sm\cite{Lab_78},
Yb\cite{KBK_91}, and
Pb\cite{DeR_83}.
For very high Rydberg states the energy spacing of states with 
different principal quantum number $n$ is small enough to cause 
hyperfine induced mixing \cite{BMT_83,SuL_88,SLB_89}.

In the present work we report on high resolution laser spectroscopic 
measurements of the hfs of $5snd\;^1\!D_2$ states in 
$^{87}$Sr and of transition isotope shifts of the stable isotopes
$^{84,86,88}$Sr for  principal quantum numbers $11\le n \le 25$.
These measurements were done with enhanced accuracy compared to 
previous results \cite{BMT_81_2,LNP_83}, in order to obtain 
sufficiently precise data for accurate comparison of the hfs with 
the results derived from second order perturbation theory.
The hfs was calculated by means of MQDT-wavefunctions in first and 
second order perturbation theory. 
We found that second order contributions need to be considered in the
analysis of the hfs in order to explain the 
unexpected large hyperfine constant $B$ and the shift of the hyperfine 
multiplet, which is responsible for large deviations from the linear 
dependence in a King Plot analysis.
\section{Experiment}
The experimental set-up used in the present work has been described in 
detail in an earlier paper \cite{HWM_88_1}. Briefly, it consisted of 
a cw narrow bandwidth ring dye laser (Spectra Physics Model 380D)
pumped by an argon ion laser (Spectra Physics Model 2045).
Typical operating conditions were uv pump powers of $2.5$--$3.0\,$W, 
resulting with a stilben 3 dye in $200$--$400\,$mW single-mode power.
The dye laser is tunable 
between $410$ and $470\,$nm. With these wavelengths Sr~I Rydberg 
levels above $n=8$ can be excited from the $5s^2\;^1\!S_0$ ground 
state by Doppler-free two-photon excitation. A Michelson-type wave 
meter (Burleigh WA 20) with $300\,$MHz accuracy was used for 
identifying Rydberg levels up to $n=200$.
For measuring the hfs -- splittings the laser frequency was stabilized
to an external confocal cavity. This, in turn, was controlled by an 
active molecular frequency standard with an accuracy of better than 
$100\,$kHz, using several precisely known $I_2$ levels combined 
with the frequency-offset-locking technique \cite{GJM_80}.
A computer (HP 21 MX-E) was employed for real-time data acquisition
and for scanning and controlling the laser frequency.

\begin{figure}
\epsfig{file=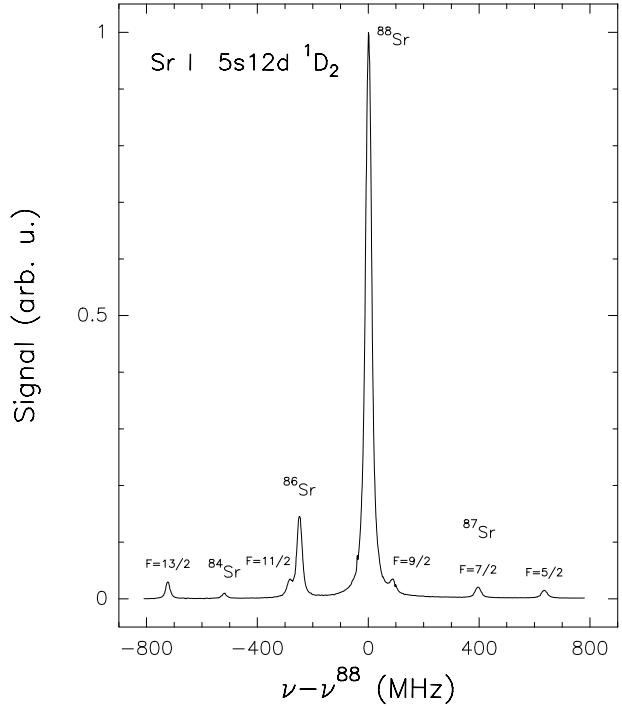, width=\columnwidth}
\protect\caption{\label{fig_spectrum}\sloppy Two-photon excitation
spectrum of the $5s^2\;^1\!S_0\rightarrow 5s12d\;^1\!D_2$ transition as
a function of the energy detuning relative to the signal of the isotope
$^{88}$Sr. }
\end{figure}

The Sr-atoms were excited and, following collisional ionization, 
detected with high efficiency in a thermionic ring diode
\cite{BMT_84}. In order to improve the signal to noise ratio, the 
retroreflected laser beam was chopped at a frequency of $80\,$Hz for 
lock-in detection (EG\&G PAR Model 5210).

A typical spectrum of the $5s12d\;^1\!D_2$ state is shown in 
Fig.\ \ref{fig_spectrum}. To determine the line positions, Lorentzian line 
profiles were fitted to the experimental data. The accuracy of the 
line distances was enhanced by recording and evaluating several 
spectra (up to 40) for each state. The mean values of the transition
frequency differences with regard to $^{88}$Sr are listed in 
Table \ref{table_pos}.
\section{Theory of hyperfine structure}
\label{sec_theory}
The hyperfine interaction describes the interaction between the 
electrons and the magnetic and electric multipole moments of the 
nucleus, where the even magnetic and the odd electric multipole 
moments vanish.
In most cases it is sufficient to take into account only the 
interaction with the nuclear magnetic dipole moment 
$\mbox{\boldmath $\mu$}_I = g_I {\bf I}$ and the electric quadrupole 
moment ${\bf Q}$.
According to Sobel'man \cite{Sob_72}, the Hamiltonian can than be 
expressed as
\begin{equation}
H_{\mbox{hfs}} = H_{\mu} + H_{Q}\;,
\end{equation}
where
\begin{align}
\begin{split}
H_{\mu} &= \sum_i [a_{{\ell}_i}({\mbox{\boldmath $\ell$}}^{(1)}
                 -\sqrt{10}[ {\bf s}^{(1)} \times 
                  {\bf C}^{(2)})]^{(1)})
\\
              & \qquad +a_{s_i}\delta_{{\ell}_i,0}{\bf
                 s}^{(1)}]\cdot
                        {\bf I}^{(1)}\quad ,
\end{split}\\
H_{Q}   &= -e^2\sum_i {r_i}^{-3}{\bf C}^{(2)}\cdot
                            {\bf Q}^{(2)}\;,
\end{align}
and ${\mbox{\boldmath $\ell$}}^{(1)}$ and ${\bf s}^{(1)}$ are the 
single electron angular momentum and spin operators, respectively. 
The tensor ${\bf C}^{(2)}$ is connected to the spherical harmonics 
$Y_{\ell m}$ by $C_m^{\ell} = [4\pi/(2\ell + 1)]^{1/2}Y_{\ell m}$.

In first order perturbation theory the hyperfine interaction yields an 
energy splitting
\begin{equation}
\Delta E_F^{(1)} = \langle \gamma JIF|H_{\mbox{hfs}}|
                           \gamma JIF\rangle
\end{equation}
of a state with total angular momentum ${\bf F}={\bf I}+{\bf J}$. Here
${\bf I}$ and ${\bf J}$ are the angular momenta of the nucleus and the
electrons, respectively.
Evaluating this matrix element, one obtains the well known Casimir 
formula \cite{Cas_63}
\begin{align}
\label{eq_casimir}
\Delta E_F^{(1)} &=  \frac{1}{2}AK +
             B\frac{\frac{3}{4}K(K+1)-J(J+1)I(I+1)}{2I(2I-1)J(2J-1)}
             \;,\\
\intertext{where}
K &= F(F+1)-J(J+1)-I(I+1)\;.
\end{align}
The hfs constants $A$ and $B$ describe the magnitudes of 
the magnetic dipole and electric quadrupole interactions, 
respectively.
Due to the different dependence on $F$ of the two terms in 
Eq.\ (\ref{eq_casimir}), it is possible to fit the Casimir formula to 
experimentally determined hfs -- splittings, and to separate the 
contributions of the magnetic dipole and the electric quadrupole 
interaction.

The measurement of the isotope shift (IS) is also possible since 
for first order perturbation theory the 
center of gravity of a hyperfine multiplet remains unshifted
($\sum_F \Delta E_F^{(1)} (2F+1) = 0$).

The situation changes if one considers contributions of second order 
perturbation theory. The second order  energy splitting of a state
$|\gamma JIF\rangle$ is
\begin{equation}
\label{eq_splitt_2}
\Delta E_F^{(2)} = \sum_{\gamma 'J'\ne \gamma J}\frac{
|\langle \gamma JIF|H_{\mbox{hfs}}|\gamma ' J'IF\rangle|^2}
{(E_{\gamma J}-E_{\gamma 'J'})}\;.
\end{equation}
Hence when second order effects are not negligible, i.e.\
when neighbouring fine structure levels with equal quantum number $F$
are lying close together, the energy 
splitting can no longer be approximated by the Casimir formula. Then 
it is not possible to separate magnetic dipole and electric quadrupole 
contributions from the experimental hfs -- splittings, because the 
second order contribution of the magnetic dipole interaction has a 
similar dependence on $F$ as the first order contribution of the 
electric quadrupole interaction. This is especially important when
the hfs is dominated by the magnetic dipole interaction.

The second order energy splitting goes together with a shift of the
center of gravity, since the contribution of each fine structure
level has a different $F$ dependence (only states with equal $F$ are 
connected).
This has the consequence that the IS of a perturbed hyperfine 
multiplet can only be extracted using second order perturbation 
theory.

To study the influence of the second order contributions, one can 
calculate the hfs in first and second order perturbation theory 
if the fine structure splitting is known. For a two electron atom like 
Strontium the magnetic dipole and electric quadrupole interaction can be 
written in terms of the one electron matrix elements 
$\langle a_{{\ell}}\rangle$ and $\langle b_{{\ell}}\rangle$.
This has been worked out by Lurio {\it et al.} \cite{LMN_62} for a 
$s\ell$-configuration.
Explicit expressions for a more general $\ell_1\ell_2$-configuration 
are given in the Appendix of the present work.
The one electron matrix elements depend on the expectation value
of $r^{-3}$ for $\ell\ne 0$ and on the probability density at the 
nucleus, $|\psi(0)|^2$, for $s$-electrons.
According to Sobel'man \cite{Sob_72} $\langle a_{{\ell}}\rangle$
and $\langle b_{{\ell}}\rangle$ can be written in units of 1 Rydberg 
($R_{\infty}$) as~:
\begin{align}
\label{eq_aln0}
\langle a_{{\ell\ne 0}}\rangle &= g_I\alpha^2\frac{m}{m_p}
        \langle R_{n\ell}|\frac{a_0^3}{r^3}|R_{n\ell}\rangle
                R_{\infty}\\
\label{eq_bln0}
\langle b_{{\ell\ne 0}}\rangle &= 2\frac{Q}{a_0^2}
        \langle R_{n\ell}|\frac{a_0^3}{r^3}|R_{n\ell}\rangle 
                R_{\infty}\\
\label{eq_al0}
\langle a_{\ell=0}\rangle &= \frac{8\pi}{3}\alpha^2g_I
                             \frac{m}{m_p}a_0^3
                             |\psi(0)|^2 R_{\infty}\\
\langle b_{\ell=0}\rangle &= 0\;.
\end{align}
Here $g_I$ is the nuclear g-factor, $\alpha$ the fine structure 
constant, $m/m_p$ the ratio of electron and proton mass, $e$ the 
elementary charge, $R_{n\ell}$ the radial part of the one-electron
wavefunction and $a_0$ the Bohr 
radius.
To a first approximation,
$\langle R_{n\ell}|a_0^3/r^3|R_{n\ell}\rangle$ and $|\psi(0)|^2$
can be calculated using 
the formula from Ref.\ \cite{Sob_72}~:
\begin{align}
\langle R_{n\ell}|\frac{a_0^3}{r^3}|R_{n\ell}\rangle &=
    \frac{1}{\ell(\ell+1)(\ell+\frac{1}{2})}\frac{Z}{{n^*}^3}
    \nonumber\\
\label{eq_r3}
 &= \frac{1}{\ell(\ell+1)(\ell+\frac{1}{2})}
    \left(\frac{\epsilon_{n\ell}}{R_{\infty}}
    \right)^{3/2}\\
\label{eq_psi0}
|\psi(0)|^2 &= \frac{1}{\pi a_0^3}\frac{Z}{{n^*}^3}
        =\frac{1}{\pi a_0^3}\left(\frac{\epsilon_{ns}}{R_{\infty}}
                            \right)^{3/2}\;,
\end{align}
where ${n^*}^3$ is the effective principal quantum number,
$\epsilon_{n\ell}$ the binding energy (in Rydbergs) of the $nl$ 
orbital, and $Z$ the effective nuclear charge.
\section{Calculation of the hyperfine splitting with MQDT 
         wavefunctions}
\label{sec_calculation}
As shown in reference \cite{Aym_84}, the MQDT is a powerful tool to
describe perturbed Rydberg series of alkaline-earth atoms.

The empirical MQDT study by Esherick \cite{Esh_77} showed that the 
$J=2$ bound-state spectrum of Strontium can be described by a five 
channel MQDT model which includes the recoupling of $^1\!D_2$ and 
$^3\!D_2$ channels and the perturbation of the $5snd\;^1\!D_2$ and 
$^3\!D_2$ series by the $4dn's\;^1\!D_2$, $4dn's\;^3\!D_2$ and 
$5pn''p\;^1\!D_2$ channels.
Here, $n'$ and $n''$ are the principal quantum numbers of the Rydberg
series converging to the $4d$ and $5p$ ionization limits.
Within the accuracy of that analysis it was found that it is not 
necessary to include the $4dn'''d\;^1\!D_2$ channel.

\begin{figure}
\epsfig{file=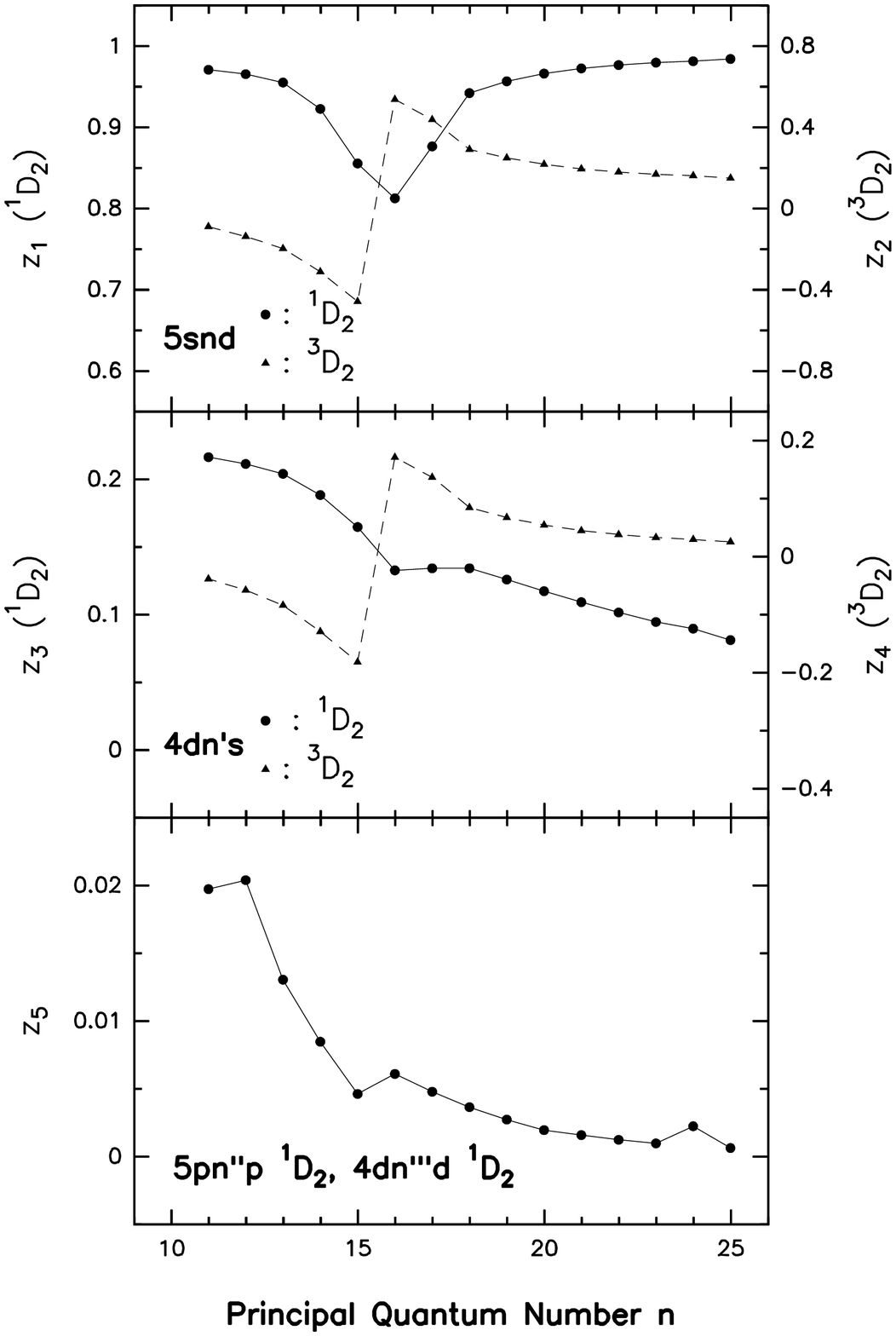, width=\columnwidth}
\protect\caption{\label{fig_misch_1d2}\sloppy Admixtures of the pure
$LS$-states according to Eq.\ \ref{eq_expand} to the {\it 5snd
$^{1}\!$D$_{2}$} Rydberg serie of Strontium calculated with the
MQDT-parameters from Ref.\ \cite{Esh_77} as a function of the principal
quantum number {\it n}.}
\end{figure}

\begin{figure}
\epsfig{file=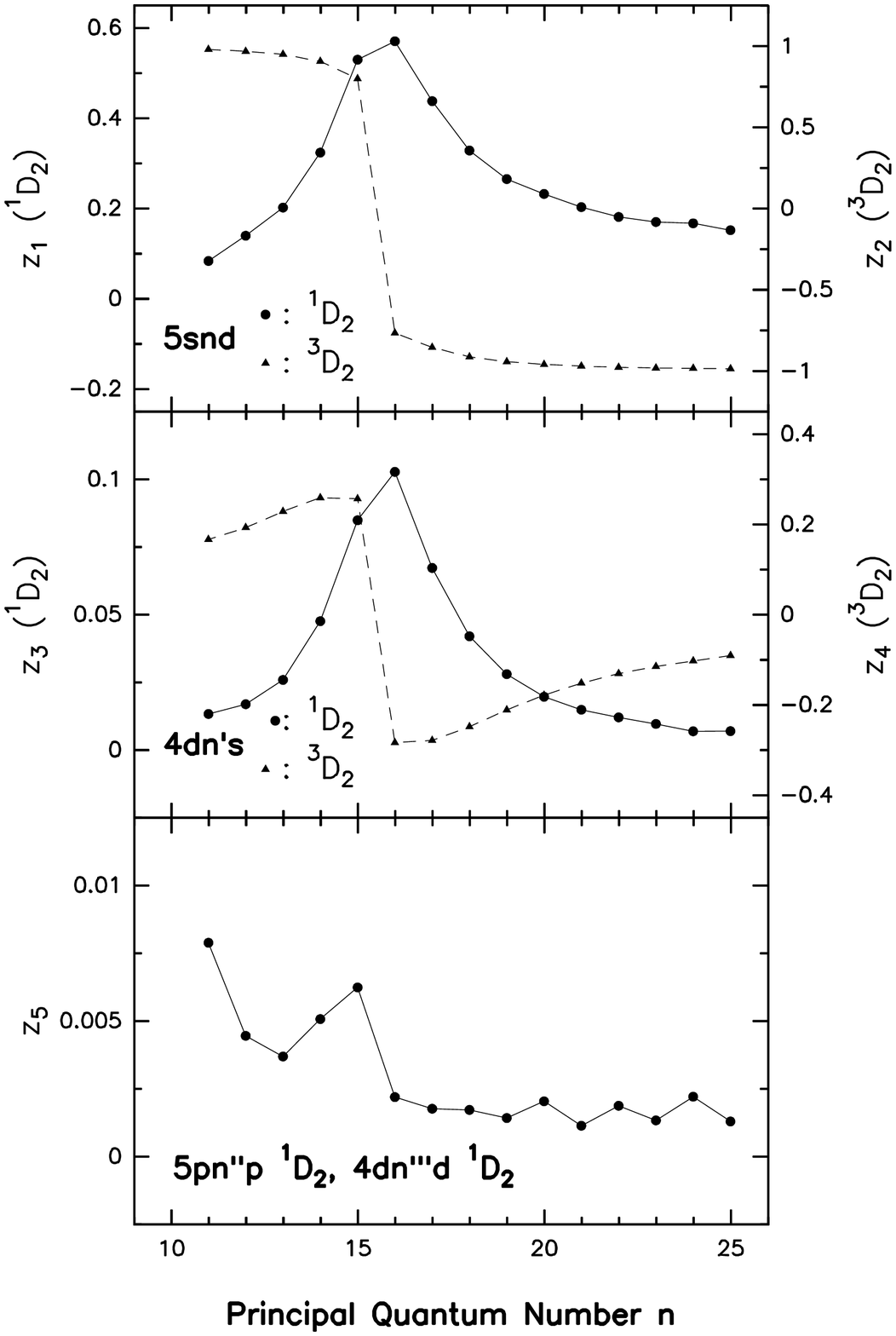, width=\columnwidth}
\protect\caption{\label{fig_misch_3d2}\sloppy Admixtures of the pure
$LS$-states according to Eq.\ \ref{eq_expand} to the {\it 5snd
$^{3}\!$D$_{2}$} Rydberg serie of Strontium calculated with the
MQDT-parameters from Ref.\ \cite{Esh_77} as a function of the principal
quantum number {\it n}.  }
\end{figure}

Since there are only a few doubly-excited states below the first 
ionization limit, the empirical MQDT study cannot give the exact 
admixture of each doubly excited channel to the levels although the 
total amount of the doubly-excited channels can be well distinguished
from the singly-excited ones.
In particular Esherick's analysis can not be used to distinguish 
between the $5pn''p\;^1\!D_2$ and $4dn'''d\;^1\!D_2$ channels.
On the other hand, MCHF calculations performed 
by Aspect {\it et al.}\ \cite{ABF_84} and {\it ab initio} MQDT 
calculations of Aymar {\it et al.}\ \cite{ALW_87} indicate that the 
$4dn'''d$ component is twice as large as the $5pn''p$ one.
Thus the wavefunction of the even $J=2$ bound states of Sr can be
written as an expansion of pure $LS$ states with the following terms
\begin{multline}\label{eq_expand}
|\gamma J=2\rangle = z_1 |5snd\;^1\!D_2\rangle_{LS}
                +z_2|5snd\;^3\!D_2\rangle_{LS}\nonumber\\
              +z_3|4dn's\;^1\!D_2\rangle_{LS}
                +z_4|4dn's\;^3\!D_2\rangle_{LS}\nonumber\\
              +z_5\Big(\sqrt{\frac{1}{3}}|5pn''p\;^1\!D_2
                                           \rangle_{LS}\\
              +\sqrt{\frac{2}{3}}|4dn'''d\;^1\!D_2
                                             \rangle_{LS}
                    \Big)\;,\nonumber
\end{multline}
where the $z_i$ are the mixing coefficients obtained by using 
Esherick's \cite{Esh_77} MQDT parameter. The mixing coefficients for 
the $5snd\;^1\!D_2$ and $5snd\;^3\!D_2$ states are drawn in 
Figs.\ \ref{fig_misch_1d2} and \ref{fig_misch_3d2} for principal 
quantum numbers $11 \le n \le 25$.
The data show a strong singlet-triplet mixing in the $5snd$ 
configuration around $n=16$. The main contribution of the doubly 
excited channels in this region of principal quantum numbers comes 
from the $4dn's$ configuration, where also a singlet-triplet mixing 
occurs.

As studied earlier \cite{BMT_81_2},
the hfs of the $5snd\;^1\!D_2$ and $^3\!D_2$
states is dominated by the Fermi-contact 
term of the magnetic dipole interaction of the $5s$ electron and 
depends strongly on the mixing coefficients $z_1$ and $z_2$.
The $n\ell$ Rydberg electron has only a weak interaction with the 
nucleus.
For pure $5snd\;^1\!D_2$ Rydberg states the hfs -- splitting should 
almost vanish and for pure $5snd\;^3\!D_2$ the dipole constant $A$ 
should be nearly $a_{5s}/12$.
Only a weak quadrupole interaction due to the small admixtures of 
doubly excited channels is expected.

In the calculation of the hyperfine splitting of the  $5snd\;^1\!D_2$
states in second order pertubation theory states of different $J$ and
equal $F$ enter the matrix elements (cf.\ Eq.\ \ref{eq_splitt_2}).
These are the fine structure levels $5snd\;^3\!D_1$, $^3\!D_2$, and
$^3\!D_3$ whereas the $5snd\;^1\!D_2$ and  $^3\!D_2$ states are 
described by Eq.\ \ref{eq_expand},
the expansion of the $5snd\;^3\!D_1$ and  $5snd\;^3\!D_3$ states 
are taken from the two channel MQDT analysis given in 
Ref.\ \cite{BeS_83} in which the mixing of the $5snd$ and $4dn's$ 
configuration have been considered.
Thus the wavefunctions of the $J=1$ and $J=3$ states can be written 
as:
\begin{align}
 |\gamma J=1\rangle 
&=
\alpha_1|5snd\,^3\!D_1\rangle_{LS}+ \alpha_2|4dn's\,^3\!D_1\rangle_{LS}\\
|\gamma J=3\rangle &= \beta_1|5snd\,^3\!D_3\rangle_{LS}+
\beta_2|4dn's\,^3\!D_3\rangle_{LS}
\end{align}
The mixing coefficients $\alpha_i$ and $\beta_i$ can be
calculated with the MQDT parameters given in Ref. \cite{BeS_83}.

The evaluation of the matrix elements is done here without 
considering matrix elements between different configurations 
$\ell_1\ell_2$.
This is a reliable procedure because these matrix elements vanish in 
most cases,
especially for the Fermi-contact term, or are very small due to the 
different effective quantum numbers.
However, in special situations \cite{JDE_81} the off-diagonal 
configuration matrix elements can be quit large.

\begin{figure}
\epsfig{file=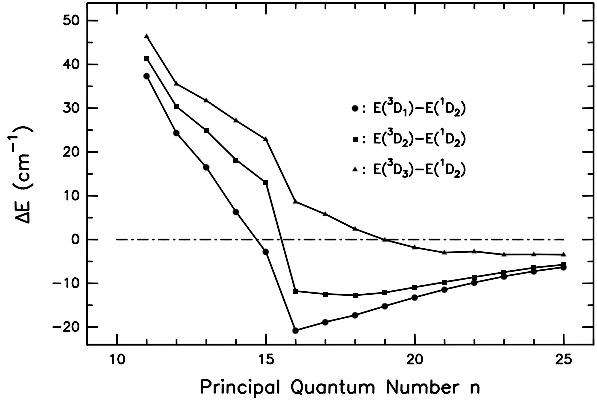, width=0.65\columnwidth, angle=270}
\protect\caption{\label{fig_energies}\sloppy {\it n} dependence of the
energy differences between the {\it 5snd $^{3}\!$D$_{1,2,3}$} and the
{\it 5snd $^{1}\!$D$_{2}$} states of Strontium.}
\end{figure}

Further the energy differences between the $5snd\;^1\!D_2$ states
and the other fine structure terms are needed for the calculations.
These are listed in Table \ref{table_energies}
and illustrated in Fig.\ \ref{fig_energies}. It can be seen that the 
energy differences of the three fine structure terms change their 
sign at different principal quantum numbers.
As studied earlier \cite{BST_82}, the small difference between the 
$5snd\;^1\!D_2$ and $5snd\;^3\!D_3$ series at $n=19$ results in a 
large shift of the respective hyperfine multiplets.

In the calculation of the one-electron matrix elements, the 
experimental value of $-1001(2)\,$MHz from Ref. \cite{BBH_83} was 
taken as $a_{5s}$-factor.
The other one-electron matrix elements can be well approximated by
introducing $a_{5s}$, and substituting Eqs.\ (\ref{eq_r3}) and 
(\ref{eq_psi0}) into Eqs.\ (\ref{eq_aln0}) to (\ref{eq_al0})~:
\begin{align}
\label{eq_aln0_2}
\langle a_{{\ell\ne 0}}\rangle &= \frac{3}{8}
  \frac{a_{5s}}{\ell(\ell+1)(\ell+\frac{1}{2})}
  \left(\frac{\epsilon_{n\ell}}{\epsilon_{5s}}\right)^{3/2}\\
\label{eq_al0_2}
\langle a_{ns}\rangle &= a_{5s}
  \left(\frac{\epsilon_{ns}}{\epsilon_{5s}}\right)^{3/2}\\
\label{eq_bln0_2}
\langle b_{{\ell\ne 0}}\rangle &= \frac{3}{4}\frac{Q}{a_0^2}
  \frac{a_{5s}}{\alpha^2 g_I \frac{m}{m_p}}
  \frac{1}{\ell(\ell + 1)(\ell + \frac{1}{2})}
  \left(\frac{\epsilon_{n\ell}}{\epsilon_{5s}}\right)^{3/2}
\end{align}

\begin{table}
\caption{Level energies of bound $5snd\;^1\!D_2$, $^3\!D_1$,
         $^3\!D_2$ and $^3\!D_3$ states of Sr~I (cf.\ 
         Fig.\ \ref{fig_energies}.
         \label{table_energies}}
\begin{tabular}{rr@{}lr@{}lr@{}lr@{}l}
\hline\hline
 & \multicolumn{2}{c}{$^1\!D_2$}
 & \multicolumn{2}{c}{$^3\!D_1$}
 & \multicolumn{2}{c}{$^3\!D_2$}
 & \multicolumn{2}{c}{$^3\!D_3$}\\
$n$ & \multicolumn{2}{c}{E (cm$^{-1}$)}
  & \multicolumn{2}{c}{E (cm$^{-1}$)}
  & \multicolumn{2}{c}{E (cm$^{-1}$)}
  & \multicolumn{2}{c}{E (cm$^{-1}$)}\\
\hline
11 & 44578.&6890$^{\rm a}$ & 44616.&05$^{\rm c}$ 
   & 44620.&08$^{\rm d}$ & 44625.&1$^{\rm e}$  \\
12 & 44829.&6648$^{\rm a}$ & 44854.&02$^{\rm b}$ 
   & 44860.&06$^{\rm b}$ & 44865.&22$^{\rm b}$ \\
13 & 45012.&0249$^{\rm a}$ & 45028.&55$^{\rm b}$ 
   & 45036.&95$^{\rm b}$ & 45043.&79$^{\rm b}$ \\
14 & 45153.&2785$^{\rm a}$ & 45159.&60$^{\rm b}$ 
   & 45171.&49$^{\rm b}$ & 45180.&44$^{\rm b}$ \\
15 & 45263.&6196$^{\rm a}$ & 45260.&84$^{\rm b}$ 
   & 45276.&65$^{\rm b}$ & 45286.&53$^{\rm b}$ \\
16 & 45362.&1272$^{\rm a}$ & 45341.&36$^{\rm b}$ 
   & 45350.&35$^{\rm b}$ & 45370.&76$^{\rm b}$ \\
17 & 45433.&2717$^{\rm a}$ & 45414.&43$^{\rm b}$ 
   & 45420.&84$^{\rm b}$ & 45439.&08$^{\rm b}$ \\
18 & 45492.&6101$^{\rm a}$ & 45475.&35$^{\rm b}$ 
   & 45479.&88$^{\rm b}$ & 45495.&02$^{\rm b}$ \\
19 & 45542.&2955$^{\rm a}$ & 45527.&10$^{\rm b}$ 
   & 45530.&18$^{\rm b}$ & 45542.&23$^{\rm b}$ \\
20 & 45584.&1831$^{\rm a}$ & 45570.&97$^{\rm b}$ 
   & 45573.&28$^{\rm b}$ & 45582.&38$^{\rm b}$ \\
21 & 45619.&7872$^{\rm a}$ & 45608.&37$^{\rm b}$ 
   & 45610.&07$^{\rm b}$ & 45616.&80$^{\rm b}$ \\
22 & 45650.&2617$^{\rm a}$ & 45640.&43$^{\rm b}$ 
   & 45641.&68$^{\rm b}$ & 45647.&54$^{\rm b}$ \\
23 & 45676.&5325$^{\rm a}$ & 45668.&11$^{\rm b}$ 
   & 45669.&06$^{\rm b}$ & 45673.&10$^{\rm b}$ \\
24 & 45699.&3308$^{\rm a}$ & 45692.&08$^{\rm b}$ 
   & 45692.&90$^{\rm b}$ & 45695.&94$^{\rm b}$ \\
25 & 45719.&2336$^{\rm a}$ & 45712.&94$^{\rm b}$ 
   & 45713.&50$^{\rm b}$ & 45715.&80$^{\rm b}$ \\
\hline\hline
\multicolumn{9}{l}{\small $^{\rm a}$Reference \cite{BLT_82_2}.}\\
\multicolumn{9}{l}{\small $^{\rm b}$Reference \cite{BLT_82_1}.}\\
\multicolumn{9}{l}{\small $^{\rm c}$Calculated value from reference
 \cite{Sch_84}.} \\
\multicolumn{9}{l}{\small $^{\rm d}$Reference \cite{Esh_77}.}\\
\multicolumn{9}{l}{\small $^{\rm e}$Reference \cite{Moo_71_2}.}
\end{tabular}
\end{table}

According to MQDT, the binding energies of the $5s$, $5p$ and $4d$ 
orbitals are assumed to be the same as those of the corresponding 
$5s$, $5p$ and $4d$ orbitals in Sr$^+$. 
The additional binding energy of the $nd$, $n's$, $n''p$ or $n'''d$ 
electron in a state can be approximately obtained by subtracting the
energy of the inner orbit (eg. $5s$) from the energy of the state 
$|\gamma J\rangle$.
The energies of the $5s$, $5p$ and $4d$ orbits in Sr$^+$ were taken 
from Ref. \cite{Moo_71_2}, where the mean values of the fine structure 
components were used for the $5p$ and $4d$ orbits.
For the quadrupole moment of $^{87}$Sr, we used the value of
$0.335(20)\cdot 10^{-28}\,$m$^2$ from Ref. \cite{HeB_77}, and for 
the dipole moment $\mu_I = g_I I = -1.089299(1)\mu_N$ reported in 
Ref. \cite{LeS_78}.
The matrix elements in Eqs.\ (\ref{eq_aln0_2}) to (\ref{eq_bln0_2}) 
were then introduced into the expressions listed in the Appendix to 
yield the energy splittings.

\begin{figure}
\epsfig{file=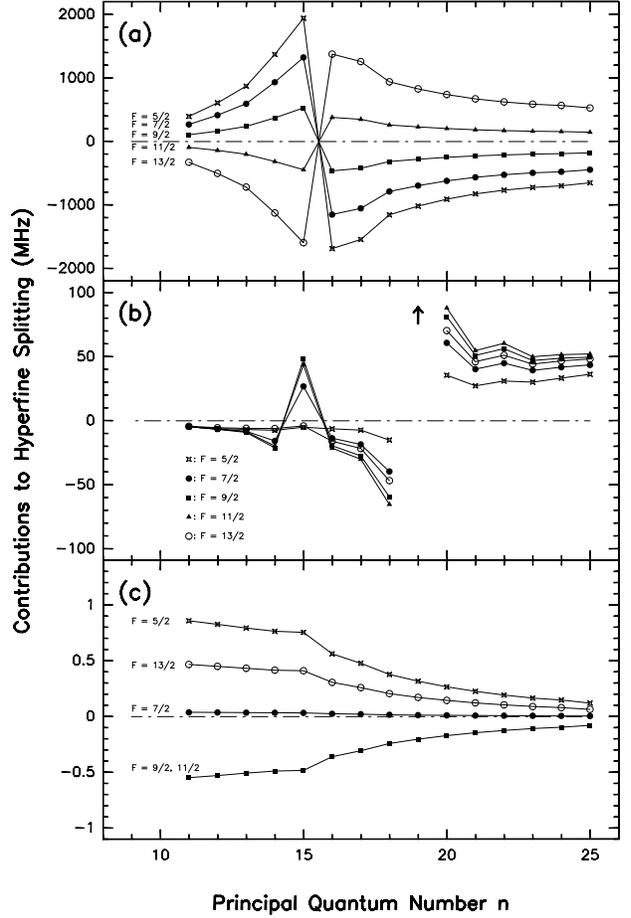, width=\columnwidth}
\protect\caption{\label{fig_contrib}\sloppy Calculated
energy-contributions to the hyperfine splitting of the {\it 5snd
$^1\!$D$_2$} states in $^{87}$Sr as a function of {\it n}.
(a)~:~magnetic dipole interaction in first order, (b)~:~magnetic dipole
interaction in second order (the very large contribution of the {\it
5s19d $^1\!$D$_2$} is only indicated by the arrow and is discussed in
detail in Ref. \cite{BST_82}), and (c)~:~electric quadrupole interaction
in first order perturbation theory.}
\end{figure}

The results of the calculations are shown in Fig.\ \ref{fig_contrib}. 
The second order contributions of the electric
quadrupole interaction are less than $1\,$kHz and will be neglected 
here.
It should be emphasized here that the first order electric quadrupole 
contributions are approximately one hundred times smaller than the 
second order magnetic dipole contributions.
The second order contribution of the magnetic dipole interaction in 
the $5s19d\;^1\!D_2$ state is not shown here because of the small 
energy separation to the $5s19d\;^3\!D_3$ state ($\approx 2\,$GHz).
In that particular case the contributions to the hfs are large in any 
higher order, which means that perturbation theory is not longer 
applicable. 
This is discussed in detail in Ref. \cite{BST_82}.

Due to the strong Fermi-contact interaction of the $5s$ electron, the 
calculated total hfs -- splittings for these states depend almost 
entirely on the mixing coefficients $z_1$ and $z_2$ and on the 
experimental value of the $a_{5s}$-factor.
\section{Comparison with the experimental data}
\subsection{Hyperfine Splitting}
\begin{figure}
\epsfig{file=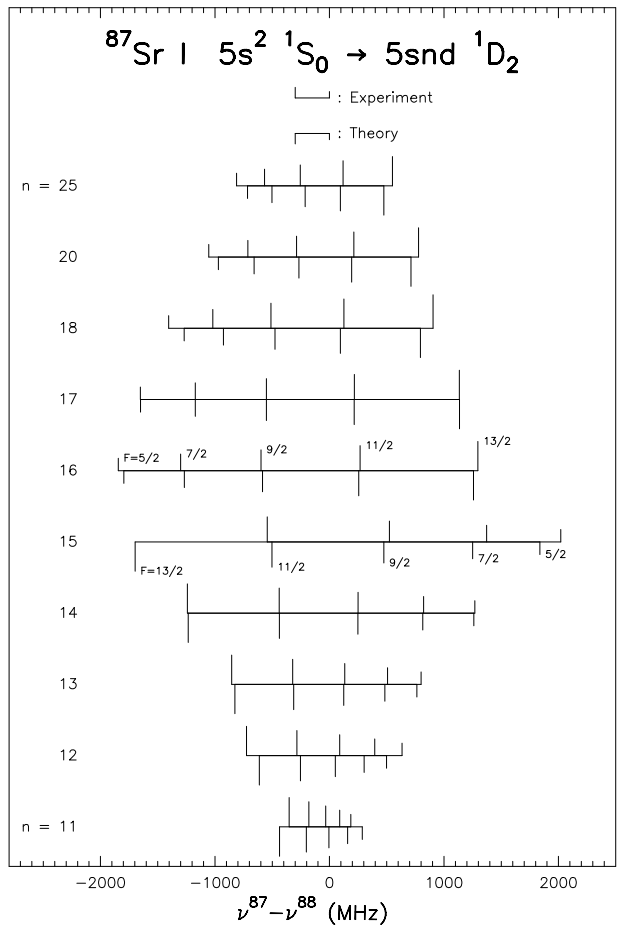, width=\columnwidth}
\protect\caption{\label{fig_pos_exptheo}\sloppy Comparison between the
measured and calculated (in second order) hyperfine splitting of the
{\it 5snd $^1\!$D$_2$} states in $^{87}$Sr.  The isotope shift is taken
into account by shifting the centers of gravity of the calculated
multiplets for agreement with the experimental ones.}
\end{figure}

\begin{figure}
\epsfig{file=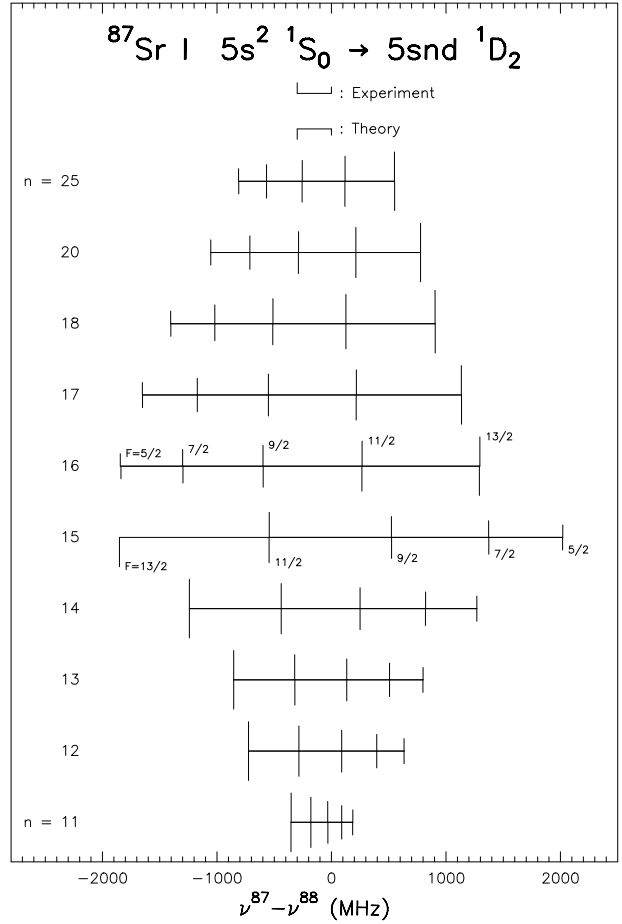, width=\columnwidth}
\protect\caption{\label{fig_posfit_exptheo}\sloppy Comparison between
the measured and calculated hyperfine splitting of the {\it 5snd
$^1\!$D$_2$} states in $^{87}$Sr.  Compared to Fig.\
\ref{fig_pos_exptheo} the theoretical line positions are optimized by a
least square fit of the mixing coefficients $z_1$, $z_2$ and the center
of gravity of the multiplet.}
\end{figure}

\begin{figure}
\epsfig{file=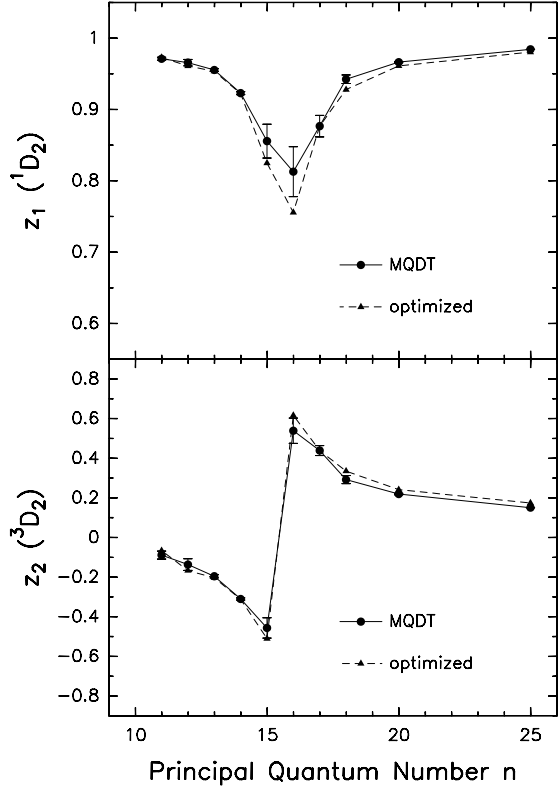, width=0.9\columnwidth}
\protect\caption{\label{fig_mix_fit}\sloppy Comparison between the
mixing coefficients $z_1$ and $z_2$ of the {\it 5snd $^1\!$D$_2$} states
in Sr calculated with the MQDT parameters from Ref.\ \cite{Esh_77} and
optimized by least square fitting with the experimental hfs --
splitting.  The errorbars estimate the uncertainty in calculating the
mixing coefficients from the MQDT parameters.}
\end{figure}

In Fig.\ \ref{fig_pos_exptheo} the calculated hfs -- splittings 
are compared with the experimental data. The isotope 
shifts of the experimental data are taken into account by shifting 
the centers of gravity of the calculated hyperfine multiplets to 
overlap with the experimental ones.
Good agreement between experimental and theoretical data is achieved.
The small deviations of the individual components
can be explained by the fact, that the hfs 
is much more sensitive to the mixing parameters in the expansion of 
Eq.\ (\ref{eq_expand}), especially to $z_1$ and $z_2$, than to the 
energies of the states which are used in the empirical MQDT analysis 
(see Ref.\ \cite{BLT_83}).
To emphasize this point we have done a least square fit of the 
theoretical line positions for each state by varying the center of 
gravity of the multiplet and the mixing coefficients $z_1$ and $z_2$ 
under the condition $z_1^2 + z_2^2 = \mbox{\it constant.}$
With only small changes of the mixing coefficients,
which cause large effects on the hfs splitting, one can obtain
a perfect agreement between experimental and theoretical line 
positions, as is shown in Fig. \ref{fig_posfit_exptheo}.
The optimized mixing coefficients differ very little from the MQDT 
mixing coefficients except near $n=15$, as can be seen in Fig. 
\ref{fig_mix_fit} and Table \ref{table_mix_fit}.
Therefore, in the following discussion it is sufficient to use the 
MQDT mixing coefficients for the description of the main features of 
the hfs splitting.

\begin{figure}
\epsfig{file=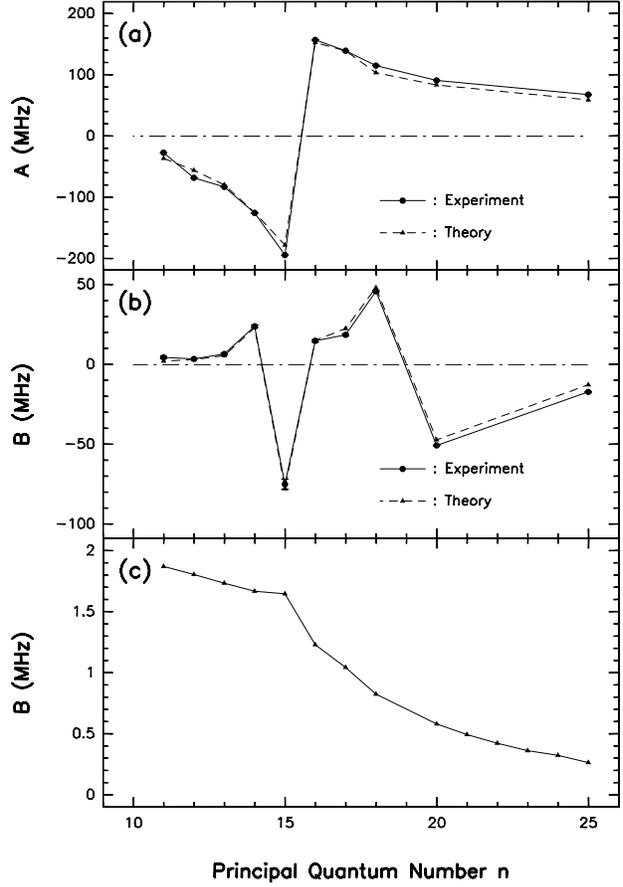, width=\columnwidth}
\protect\caption{\label{fig_ab}\sloppy Hyperfine structure constants $A$
and $B$ for the {\it 5snd $^1\!$D$_2$} states in $^{87}$Sr as a function
of {\it n} (see Table \ref{table_ab}).  In (a) and (b): $A$ and $B$ are
determined by fitting the Casimir formula to the measured and calculated
line positions.  (c): Theoretical prediction for $B$ in first order.  It
should be noted that the constant $B$ in (b) does not reflect the amount
of the electric quadrupole interaction but rather the second order
magnetic contribution.  }
\end{figure}

The need for considering second order effects 
for the hfs becomes evident, 
when one tries to separate the magnetic dipole and electric 
quadrupole contributions by fitting the Casimir formula 
(which contains only first order effects)
to the measured and calculated hfs -- splittings.
This is shown in Fig.\ \ref{fig_ab}(a) and \ref{fig_ab}(b)
and is listed in Table \ref{table_ab}.
At the first glance an excellent agreement between 
experiment and theory is obtained. However, the large 
constant $B$ in Fig.\ \ref{fig_ab}(b) does not describe the electric 
quadrupole interaction, and reflects mainly the second order 
contribution of the magnetic dipole interaction.
The contribution of the quadrupole interaction is very small indeed,
and is determined by the admixture of the $4dn's$ configuration.
This is shown in Fig.\ \ref{fig_ab}(c), where the
electric quadrupole interaction is considered to first order only. 

\subsection{Isotope Shift}
\begin{figure}
\epsfig{file=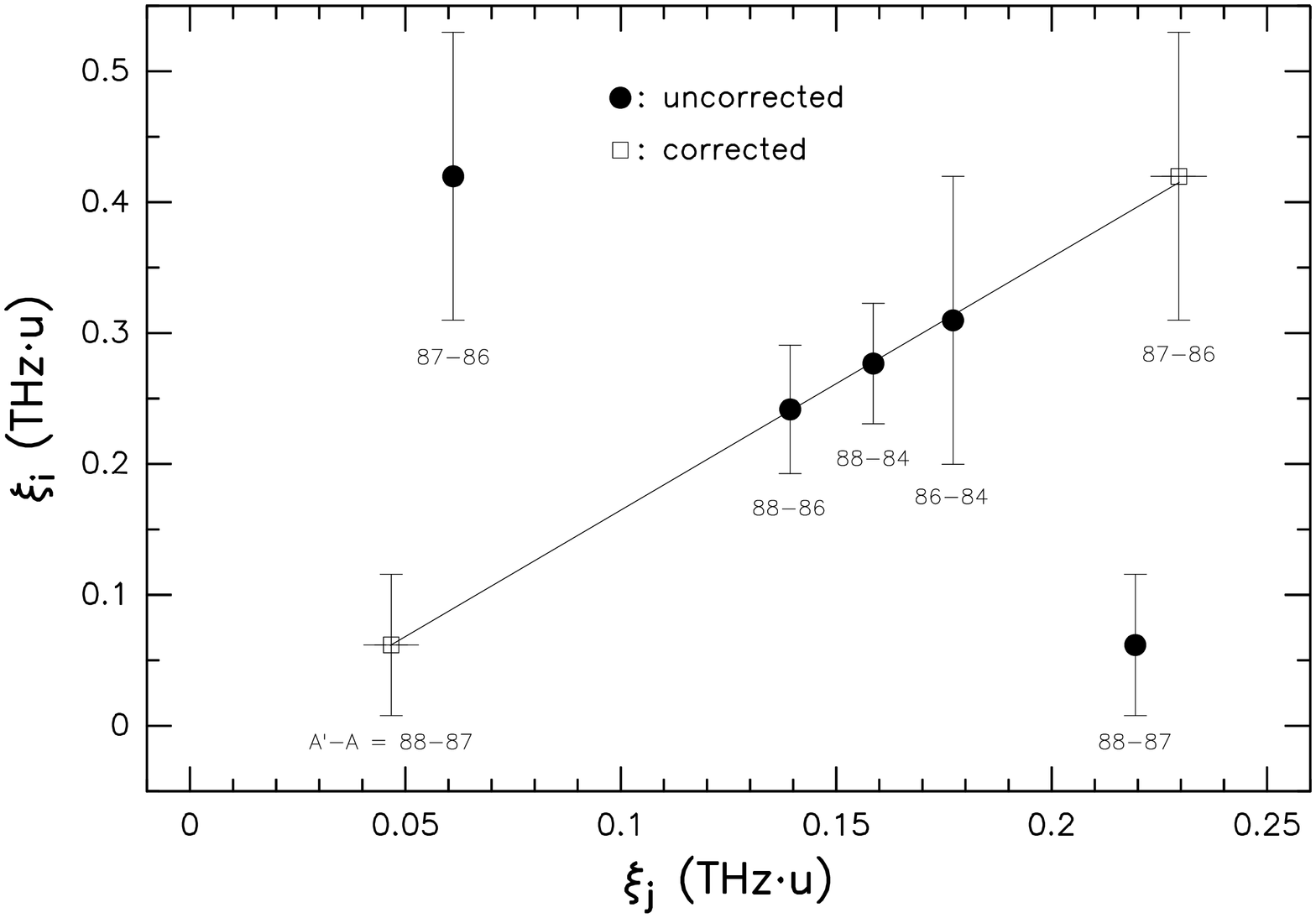, width=0.65\columnwidth, angle=270}
\protect\caption{\label{fig_king}\sloppy King plot for the two-photon
transition j: Sr~I $5s^2\;^1\!S_0\rightarrow 5s17d\;^1\!D_2$.  The
values for the reference transition i: Sr~II $5s\;^2\!S_{1/2}\rightarrow
5p\;^2\!P_{1/2}$ are taken from Ref.\ \cite{BBH_83}.  The shifts for the
isotope pairs involving $^{87}$Sr are corrected by an amount given by
the center of gravity of the theoretical hyperfine splitting listed in
Table \ref{table_ab}.}
\end{figure}

The influence of the second order effects on the IS can be 
seen in a King-plot analysis in Fig.\ \ref{fig_king}. The King-plot 
procedure (see Ref.\ \cite{HeS_74}) can be used for testing the 
consistency of the IS measurement in two transitions. It allows the 
separation of the specific mass and the field shift of a transition, 
provided the specific mass and the field shift are known for a 
reference transition.
In a King-plot, the modified IS $\xi_i^{A,A'}$ of a transition
$i$ between two isotopes with mass numbers $A$ and $A'$ 
(where $A'>A$) is plotted against the modified IS $\xi_j^{A,A'}$ of
the reference transition.
The modified IS is defined as the difference between the 
experimental IS $\delta\nu_i^{A,A'}$ and the normal mass shift
multiplied by the factor $m_A m_{A'}/(m_{A'}-m_A)$, where $m_A$ and
$m_{A'}$ are the masses of the isotopes~:
\begin{equation}
\xi_i^{A,A'} = \delta\nu_i^{A,A'}\frac{m_Am_{A'}}{m_{A'}-m_A} - 
               m_e\nu_i\;.
\end{equation}
Here $m_e$ is the electron mass and $\nu_i$ the transition energy.
In a King-plot the points for different isotope pairs should lie on a 
straight line.

\begin{table}
\caption{\label{table_mix_fit} Mixing coefficients $z_1$ and $z_2$ for
the $5snd\;^{1}\!D_{2}$ Rydberg serie of Sr~I (see Eq.\
(\ref{eq_expand})) (a) calculated with the MQDT-parameters from
Reference \cite{Esh_77} and (b) optimized by least square fitting of the
theoretical line positions (see also Fig.\ \ref{fig_mix_fit}.  }
\begin{tabular}[t]{rr@{}lr@{}lr@{}lr@{}l}
\hline\hline
 & \multicolumn{4}{c}{(a)}
 & \multicolumn{4}{c}{(b)}\\
\cline{2-5}\cline{6-9}
$n$ 
 & \multicolumn{2}{c}{$z_1$} 
 & \multicolumn{2}{c}{$z_2$}
 & \multicolumn{2}{c}{$z_1$}
 & \multicolumn{2}{c}{$z_2$}\\
\hline
11 & 0.&971(3) & -0.&09(2) & 0.&9732 & -0.&0651\\
12 & 0.&966(5) & -0.&14(3) & 0.&9613 & -0.&1658\\
13 & 0.&956(2) & -0.&195(9)& 0.&9540 & -0.&2024\\
14 & 0.&923(2) & -0.&309(6)& 0.&9226 & -0.&3111\\
15 & 0.&86(3)  & -0.&46(5) & 0.&8252 & -0.&5090\\
16 & 0.&81(4)  &  0.&54(7) & 0.&7560 &  0.&6174\\
17 & 0.&88(2)  &  0.&44(3) & 0.&8771 &  0.&4400\\
18 & 0.&943(6) &  0.&29(2) & 0.&9282 &  0.&3361\\
20 & 0.&9667(9)&  0.&220(4)& 0.&9614 &  0.&2426\\
25 & 0.&9847(2)&  0.&152(1)& 0.&9808 &  0.&1753\\
\hline\hline
\end{tabular}
\end{table}

For the transitions $5s^2\;^1\!S_0\rightarrow 5snd\;^1\!D_2$ in
Sr~I this is true only for the even isotope pairs.
As an example, Fig.\ \ref{fig_king} shows a King-plot for the 
transition Sr~I $5s^2\;^1\!S_0\rightarrow 5s17d\;^1\!D_2$. The 
transition Sr~II $5s\;^2\!S_{1/2}\rightarrow 5p\;^2\!P_{1/2}$ from 
Ref.\ \cite{BBH_83} was taken as reference transition.
Obviously the points involving the odd isotope
$^{87}$Sr (which are calculated using the center of gravity of the 
experimental hfs -- splitting listed in Table \ref{table_ab}) are 
far off the linear dependence of the even isotope pairs.
This effect was also noticed in Ref.\ \cite{KBK_91} for Ytterbium, 
where the deviation from a straight line was taken to determine the 
magnitude of the second order contributions.
If one corrects the experimental IS of the $^{87}$Sr for the second
order contribution to the shift, represented by the center of 
gravity of the theoretical hfs -- splitting listed in Table 
\ref{table_ab} (which is zero without second order effects), the 
data fit a straight line with excellent agreement.
\section{Summary}
Higher order contributions to the hfs and to the IS have been studied
for $^{87}$Sr $5snd\;^1\!D_2$ states in the range  $11\le n \le 25$.
An enhanced accuracy, compared to previous experiments, made it 
possible to identify second order magnetic dipole contributions to 
the hyperfine splitting and shift.

In parallel the hfs was calculated in second order pertubation 
theory using MQDT wavefunctions. We demonstrate that the
knowledgement of second order contributions is important for 
the determination of the quadrupole splitting and the IS, 
{\it even} when the fine structure separation is large compared to 
the hfs -- splitting.
Neglecting the second order contributions in the hfs 
analysis of the experimental data leads to unrealistically large 
quadrupole factors in the Casimir formula as well as to erroneous 
positions of the centers of gravity which do not give the correct IS.

In the investigated region of $n$ of the $5snd\;^1\!D_2$ states, it 
was shown that the second order magnetic dipole contribution to the
hfs -- splitting is near $n=15$ up to one hundred times larger than 
the first order electric quadrupole contribution which is determined 
by small admixtures of doubly excited states.
Because of the dominance of the second order magnetic dipole 
contributions it was not possible to make any reliable predictions,
based on experimental data, about the admixtures 
of doubly excited channels into the wavefunction of the 
$5snd\;^1\!D_2$ states that could significantly improve
the results of a MQDT-analysis.
Evaluating the IS from the $^{87}$Sr spectrum, corrected for the 
second order contributions of the hfs to the shift, leads to a 
value consistent with a King-plot analysis.
\section{Acknowledgements}
We like to thank K.-D.~Heber and M.~A.~Khan for helpful discussions 
and Jun-qiang Sun for his help related to the calculation of the 
matrix elements.
\onecolumn
\section{Tables}
\label{tab}

\begin{table}[h]\footnotesize
\caption{Mean values of the frequency differences 
         with regard to $^{88}$Sr
         for the transitions $5s^2\;^1\!S_0\rightarrow 5snd\;^1\!D_2$ 
         of the isotope pairs ($^{87}$Sr,$^{88}$Sr), 
         ($^{86}$Sr,$^{88}$Sr) and ($^{84}$Sr,$^{88}$Sr).
        }
\begin{tabular}[t]{rr@{.}lr@{.}lr@{.}lr@{.}lr@{.}lr@{.}lr@{.}l}
\hline\hline
  & \multicolumn{10}{c}{$\nu^{87}-\nu^{88}$ (MHz)}
  & \multicolumn{2}{c}{$\nu^{86}-\nu^{88}$ (MHz)}
  & \multicolumn{2}{c}{$\nu^{84}-\nu^{88}$ (MHz)}\\  
\cline{2-11}
$n$ & \multicolumn{2}{c}{$F=5/2$} 
  & \multicolumn{2}{c}{$F=7/2$} 
  & \multicolumn{2}{c}{$F=9/2$}
  & \multicolumn{2}{c}{$F=11/2$} 
  & \multicolumn{2}{c}{$F=13/2$} 
  & \multicolumn{4}{c}{ }\\
\hline
11 & 185&83(45) & 89&81(25) & -33&67(65) & -179&95(28) 
   & -352&37(29) & -244&99(21) & -511&50(49)\\ 
12 & 634&105(95) & 395&214(82) & 88&69(12) & -285&218(84) 
   & -724&297(87) & -248&785(57) & -521&06(13)\\
13 & 799&25(20) & 507&14(42) & 133&57(28) & -320&94(25) 
   & -854&17(48) & -240&73(11) & -502&95(26)\\
14 & 1268&92(67) & 821&63(71) & 249&24(23) & -439&12(24) 
   & -1240&48(72) & -241&03(18) & 504&66(30)\\
15 & 2019&3(11) & 1372&1(14) & 523&41(80) & -543&91(73) 
   & \multicolumn{2}{c}{-}  & -246&54(83) & -516&71(68)\\
16 & -1844&34(65) & -1299&56(56) & -598&83(26) & 267&47(20) 
   & 1295&54(66) & -242&60(18) & -508&06(30)\\
17 & -1651&93(67) & -1172&44(47) & -550&56(21) & 217&64(27) 
   & 1134&18(39) & -234&36(14) & -490&29(15)\\
18 & -1404&38(42) & -1019&48(50) & -511&16(32) & 126&10(17) 
   & 904&67(26) & -230&05(15) & -481&55(24)\\
19 & \multicolumn{2}{c}{-} & \multicolumn{2}{c}{-} 
   & \multicolumn{2}{c}{-} & \multicolumn{2}{c}{-} 
   & \multicolumn{2}{c}{-} & -227&97(25) & -477&05(42)\\
20 & -1054&77(18) & -712&84(23) & -287&75(15) & 212&58(14) 
   & 777&65(18) & -226&23(14) & -472&68(29)\\
25 & -811&90(20) & -566&69(23) & -255&836(82) & 118&212(64) 
   & 550&13(21) 
   & -222&179(49) & -464&76(18)\\
\hline
\end{tabular}
\label{table_pos}
\end{table}

\begin{table}
\caption{\label{table_ab} $A$ and $B$ constants and centers of gravity
         for the $5snd\;^1\!D_2$ states of $^{87}$Sr determined by
         fitting the Casimir formula to the measured and calculated line
         positions (see Fig.\ \ref{fig_ab}(a) and \ref{fig_ab}(b).  The
         theoretical line positions are calculated with the mixing
         coefficients obtained from the MQDT parameters given in Ref.\
         \cite{Esh_77}.  It should be noted that the isotope shift is
         not taken into account for the theoretical values
         $\langle\Delta\nu^{87}_F\rangle$.  }
\begin{tabular}{rr@{}lr@{}lr@{}lr@{}lr@{}lr@{}l}
\hline\hline
 & \multicolumn{6}{c}{Experiment}
 & \multicolumn{6}{c}{Theory}\\
\cline{2-7}\cline{8-13}
$n$ & \multicolumn{2}{c}{$A$ (MHz)}
  & \multicolumn{2}{c}{$B$ (MHz)}
  & \multicolumn{2}{c}{$\langle \nu^{87}_F-\nu^{88}\rangle$ (MHz)}
  & \multicolumn{2}{c}{$A$ (MHz)}
  & \multicolumn{2}{c}{$B$ (MHz)}
  & \multicolumn{2}{c}{$\langle \Delta\nu^{87}_F\rangle$ (MHz)}\\
\hline
 11 & -26.&855$\;$(20)&  4.&49$\;$(57)&-111.&72$\;$(15)& -36.&06
    &  2.&2& -4.&48\\
 12 & -67.&895$\;$(57)&  3.&55$\;$(15)&-114.&21$\;$(10)& -55.&50
    &  3.&2& -6.&11\\
 13 & -82.&605$\;$(26)&  6.&59$\;$(50)&-112.&43$\;$(16)& -79.&39
    &  5.&6& -7.&56\\
 14 &-125.&217$\;$(37)& 24.&01$\;$(79)&-119.&22$\;$(24)&-124.&50
    & 23.&1&-13.&96\\
 15 &-194.&02$\;$(18) &-75.&0$\;$(32) & -80.&70$\;$(99)&-177.&56
    &-77.&4& 22.&74\\
 16 & 157.&248$\;$(34)& 14.&85$\;$(75)&-121.&80$\;$(21)& 152.&84
    & 15.&2&-16.&36\\
 17 & 139.&542$\;$(28)& 18.&64$\;$(61)&-126.&13$\;$(17)& 139.&57
    & 22.&5&-22.&54\\
 18 & 115.&29$\;$(13) & 45.&96$\;$(43)&-150.&29$\;$(13)& 103.&66
    & 48.&2&-48.&68\\
 20 &  91.&067$\;$(11)&-50.&62$\;$(25)& -29.&50$\;$(10)&  83.&65
    &-47.&1& 71.&22\\
 25 &  67.&954$\;$(69)&-17.&12$\;$(24)& -56.&78$\;$(10)&  59.&36
    &-12.&6& 47.&38\\
\hline\hline
\end{tabular}
\end{table}

\onecolumn
\section{Hyperfine interaction matrix elements}
\label{appendix}
\subsection{Magnetic dipole interaction}
\begin{multline}
\langle {\ell}_1 {\ell}_2 SLJIF|
           H_\mu
         |{\ell}_1 {\ell}_2 S'L'J'IF\rangle
= (-1)^{J'+I+F}\sqrt{I(I+1)(2I+1)}\left\{\begin{array}{ccc}
                                         J & I & F\\
                                         I & J'& 1
                                  \end{array}\right\}\nonumber\\
     \times({\ell}_1 {\ell}_2 SLJ||
                \sum_{i=1}^2 a_{{\ell}_i}
                            ({\mbox{\boldmath $\ell$}}^{(1)}
              - \sqrt{10}[ {\bf s}^{(1)} \times {\bf C}^{(2)})]^{(1)}
              + a_{s_i}\delta_{{\ell}_i,0}{\bf s}^{(1)}
                                    ||{\ell}_1 {\ell}_2 S'L'J')\;,
\end{multline}
\begin{multline}
({\ell}_1 {\ell}_2 SLJ||
     \sum_{i=1}^2 a_{{\ell}_i}{\mbox{\boldmath $\ell$}}^{(1)}
                      ||{\ell}_1 {\ell}_2 S'L'J')\\
=  (-1)^{S+L'+J+{\ell}_1+{\ell}_2}\sqrt{(2J+1)(2J'+1)(2L+1)(2L'+1)}
                                  \left\{\begin{array}{ccc}
                                         L & J & S\\
                                         J' & L'& 1
                                  \end{array}\right\}\\
   \times \bigg[ \langle a_{{\ell}_1}\rangle
               (-1)^{L'}\sqrt{{\ell}_1({\ell}_1+1)(2{\ell}_1+1)}
                        \left\{\begin{array}{ccc}
                               {\ell}_1 &  L  & {\ell}_2\\
                               L'  & {\ell}_1 & 1
                               \end{array}\right\}\\ 
  \quad   +\langle a_{{\ell}_2}\rangle
               (-1)^{L}\sqrt{{\ell}_2({\ell}_2+1)(2{\ell}_2+1)}
                        \left\{\begin{array}{ccc}
                               {\ell}_2 &  L  & {\ell}_1\\
                               L'  & {\ell}_2 & 1
                               \end{array}\right\}\bigg]\delta_{S,S'}\;,
\end{multline}
\begin{multline}
\lefteqn{({\ell}_1 {\ell}_2 SLJ||
            \sum_{i=1}^2 a_{{\ell}_i}[{\bf s}^{(1)}\times
                                      {\bf C}^{(2)}]^{(1)}
                               ||{\ell}_1 {\ell}_2 S'L'J')}\\
= (-1)^{{\ell}_1+{\ell}_2}\frac{3}{\sqrt{2}}
      \sqrt{(2J+1)(2J'+1)(2L+1)(2L'+1)(2S+1)(2S'+1)}
      \left\{\begin{array}{ccc}
              {\scriptstyle \frac{1}{2}} & S & {\scriptstyle 
                            \frac{1}{2}}\\
               S'    & {\scriptstyle \frac{1}{2}} & 1
             \end{array}\right\}
       \left\{\begin{array}{ccc}
                           S & S'& 1\\
                           L & L'& 2\\
                           J & J'& 1
             \end{array}\right\}\\
    \times \bigg[ \langle a_{{\ell}_1}\rangle
               (-1)^{S'+L'}({\ell}_1||{\bf C}^{(2)}||{\ell}_1)
               \left\{\begin{array}{ccc}
                                   {\ell}_1 & L   & {\ell}_2\\
                                   L'  & {\ell}_1 & 2
                      \end{array}\right\}
            +\langle a_{{\ell}_2}\rangle
               (-1)^{S+L}({\ell}_2||{\bf C}^{(2)}||{\ell}_2)
               \left\{\begin{array}{ccc}
                                   {\ell}_2 & L   & {\ell}_1\\
                                   L'  & {\ell}_2 & 2
                      \end{array}\right\}\bigg]\;,
\end{multline}
\begin{equation}
(\ell||{\bf C}^{(2)}||\ell) = - \sqrt{\frac{\ell(\ell+1)(2\ell+1)}
                                       {(2\ell+3)(2\ell-1)}}\;,
\end{equation}
\begin{multline}
({\ell}_1 {\ell}_2 SLJ||\sum_{i=1}^2 a_{s_i}\delta_{{\ell}_i,0}
                        {\bf s}^{(1)}||{\ell}_1 {\ell}_2 S'L'J')
=  (-1)^{S+L+J'+1}\sqrt{\frac{3}{2}}\sqrt{(2J+1)(2J'+1)(2S+1)
                                             (2S'+1)}\\
\times\left\{\begin{array}{ccc}
                           S & J & L\\
                           J'& S'& 1
             \end{array}\right\}      
      \left\{\begin{array}{ccc}
              {\scriptstyle \frac{1}{2}} & S & 
                       {\scriptstyle \frac{1}{2}}\\
               S'    & {\scriptstyle \frac{1}{2}} & 1
             \end{array}\right\}
             [ (-1)^{S'} \langle a_{s_1}\rangle\delta_{{\ell}_1,0}
              +(-1)^{S } \langle a_{s_2}\rangle\delta_{{\ell}_2,0}
             ]\delta_{L,L'}\;.
\end{multline}
\subsection{Electric quadrupole interaction}
\begin{multline}
\langle {\ell}_1 {\ell}_2 SLJIF|
          H_Q
         |{\ell}_1 {\ell}_2 S'L'J'IF\rangle\\
= (-1)^{J'+I+F+1}\sqrt{\frac{(2I+3)(2I+1)(I+1)}{4I(2I-1)}}
                           \left\{\begin{array}{ccc}
                                         J & I & F\\
                                         I & J'& 2
                                  \end{array}\right\}
        ({\ell}_1 {\ell}_2 SLJ||
             e^2Q\sum_{i=1}^2 r_i^{-3}{\bf C}^{(2)}
                              ||{\ell}_1 {\ell}_2 S'L'J')\;,
\end{multline}
\begin{multline}
({\ell}_1 {\ell}_2 SLJ||
             e^2Q\sum_{i=1}^2 {r_i}^{-3}{\bf C}^{(2)}
                               ||{\ell}_1 {\ell}_2 S'L'J')\\
=  (-1)^{S+L'+J+{\ell}_1+{\ell}_2}\sqrt{(2J+1)(2J'+1)(2L+1)(2L'+1)}
      \left\{\begin{array}{ccc}
                           L & J & S\\
                           J'& L' & 2
             \end{array}\right\}\\ 
\times \bigg[ \langle b_{{\ell}_1}\rangle
               (-1)^{L'}({\ell}_1||{\bf C}^{(2)}||{\ell}_1)
               \left\{\begin{array}{ccc}
                                    {\ell}_1 & L   & {\ell}_2\\
                                    L'  & {\ell}_1 & 2
                      \end{array}\right\}
              +\langle b_{{\ell}_2}\rangle
               (-1)^L({\ell}_2||{\bf C}^{(2)}||{\ell}_2)
               \left\{\begin{array}{ccc}
                                    {\ell}_2 & L   & {\ell}_1\\
                                    L'  & {\ell}_2 & 2
                      \end{array}\right\}\bigg]\delta_{S,S'}\;.
\end{multline}
\end{document}